\documentclass[12pt,iopart,a4paper]{article}
\usepackage{amssymb}
\usepackage{amsfonts}
\usepackage{amsmath}
\usepackage{cite}
\usepackage{graphicx}

\input{tcilatex}
\begin{document}

\author{P. G. C. Almeida, M. S. Benilov, and D. F. N. Santos \\
CCCEE, Universidade da Madeira, Largo do Munic\'{\i}pio,\\
9000 Funchal, Portugal\smallskip}
\title{Modelling self-organization in DC glow microdischarges: new 3D modes}
\date{}
\maketitle

\begin{abstract}
Seven new 3D modes of self-organization in DC glow discharges are computed
in the framework of the simplest self-consistent model of glow discharge.
Some of the modes branch off from and rejoin the 1D mode, while others
bifurcate from a 2D or a 3D mode. The patterns associated with computed 3D
modes are similar to patterns observed in the experiment. The computed
transition from a spot pattern comprising five spots into a pattern
comprising a ring spot also was observed in the experiment.
\end{abstract}

\section{Introduction}

Self-organization in DC glow microdischarges has been observed for the first
time a decade ago \cite{Schoenbach2004,Moselhy2004} and represents a very
interesting and potentially important phenomenon. Since then, a number of
experimental reports on this phenomenon have been published \cite%
{Takano2006,Takano2006_ICOPS,Lee2007,Zhu2007,Lee2008,2014c,Zhu2014}, as well
as a theoretical interpretation in terms of mutiple solutions existing in
the theory of glow discharge \cite{2007c,2009f,2010a,2011a,2013g,2014c}; see
also \cite{2014b} where a detailed discussion can be found.

This paper is concerned with computing some 3D modes of self-organization
which have observed in the experiment but not in the modelling. This
includes modes associated with two to six spots observed, e.g., in \cite%
{Takano2006,2014c}, and a transition from a 3D mode with several spots into
a 2D with a ring spot mode that has been observed in \cite{Zhu2014} and
suggests the existence of the corresponding bifurcation.

\section{Results}

The numerical model is identical to that described in \cite{2010a} and
successfully used in \cite{2014c} for providing a guide for experiments in
krypton microdischarges. Modelling is performed for a cylindrical discharge
vessel with interelectrode gap of $0.5\unit{mm}$ and radius of $0.5\unit{mm}$
for xenon at $30\unit{Torr}$; some results for krypton at $100\unit{Torr}$
and the same geometry are also given for comparison.

Figure 1 depicts the CVC of the 1D mode and of the first five multimensional
modes. (Here $\left\langle j\right\rangle $ is the average current density
evaluated over a cross section of the discharge vessel.) The schematics in
this figure illustrate distributions of current density on the cathode
surface associated with each mode. $a_{i}$ and $b_{i}$ designate bifurcation
points where the corresponding solution branches off from and rejoins the 1D
mode. The modes are ordered by decreasing separation of the bifurcation
points: the first mode is designated $a_{1}b_{1}$ and is the one possessing
bifurcation points further apart, the second mode is designated $a_{2}b_{2}$
and is the one possessing the second largest separation between bifurcation
points, and so on. The modes $a_{1}b_{1}$ and $a_{3}b_{3}$ have been
computed before \cite{2011a,2010a} and are included in figure 1 for the sake
of completeness; note that $a_{3}b_{3}$ is a 2D mode with one branch
associated with a spot at the centre of the cathode and the other with a
ring spot at the periphery of the cathode.

The modes $a_{2}b_{2}$, $a_{4}b_{4}$ and $a_{5}b_{5}$ are new. It is
interesting to note a retrograde section in the CVC of the mode $a_{2}b_{2}$%
, which is seen in figure 1b in a narrow current range around $280\unit{A}%
\unit{m}^{-2}$. The evolution with $\left\langle j\right\rangle $ of the
cathodic spot patterns associated with the each mode is shown in figure 2.
Let us consider first the evolution of patterns associated with the mode $%
a_{2}b_{2}$; figure 2a). The state $151.05\unit{V}$ is positioned in the
vicinity of the bifurcation point $a_{2}$ and the spot pattern comprises two
very diffuse cold spots at the periphery. Further away from $a_{2}$, the
cold phase expands and at state $151.79\unit{V}$ start merging. This is
acompanied by the above-mentioned retrograde section seen in figure 1b. As
current is further reduced towards $b_{2}$, the two cold spots expand
further and the resulting pattern comprises two well-pronounced hot spots at
the periphery; state $160.4\unit{V}$. The state $173.93\unit{V}$ is
positioned in the vicinity of the bifurcation point $b_{2}$ and the hot
spots are very diffuse. Patterns with two spots similar to that of the state 
$160.4\unit{V}$ have been observed in the experiment \cite{Takano2006}. A
difference between the pattern of the state $160.4\unit{V}$ and the
experimentally observed pattern is that in the modelling the spots are
positioned at the periphery and not inside the cathode, however this
difference will disappear if neutralization of charged particles at the wall
of the discharge vessel is taken into account \cite{2013g}.

The patterns associated with the mode $a_{4}b_{4}$ are shown in figure 2b).
The state $151.01\unit{V}$ is positioned in the vicinity of the bifurcation
point $a_{4}$ and the pattern is very diffuse. Further away from $a_{4}$,
the spots become better pronounced and a cold spot appears at the centre. As
current is further reduced towards $b_{4}$, the cold spot at the centre is
gradually tranformed into a hot spot. Eventually, the hot spots become well
pronounced and a pattern comprising three (hot) spots at the periphery and a
central spot is formed. (It is this pattern which is shown in figure 1.) The
state $172.48\unit{V}$ is positioned in the vicinity of the bifurcation
point $b_{4}$ and the hot spots are very diffuse. The transition between
patterns with well-defined cold and hot spots is not accompanied by
retrograde behaviour, in contrast to the case of the mode $a_{2}b_{2}$.
Patterns with three spots similar to that of the state $151.53\unit{V}$ have
been observed in the experiment \cite{Takano2006,Zhu2014}; patterns with
three spots at the periphery and a spot at the centre similar to that of the
state $151.74\unit{V}$ have also been observed in the experiment \cite%
{Lee2008}.

The evolution of patterns associated with the mode $a_{5}b_{5}$ shown in
figure 2c) follows the same trend as the mode $a_{4}b_{4}$.

Figure 3 depicts mode $a_{2}b_{2}$ computed for krypton and $100\unit{Torr}$%
. The trend is the same that was found in xenon, except that the retrograde
section associated with the switching between cold and hot spots is much
wider in krypton and spans a range of about $300\unit{A}\unit{m}^{-2}$.

A convenient graphic representation of the modes $a_{4}b_{4}$ and $%
a_{5}b_{5} $ is given in figure 4 with the use of the coordinates $\left(
j_{c},\left\langle j\right\rangle \right) $, where $j_{c}$ is the current
density at the centre of the cathode. Also shown for the sake of
completeness are the modes $a_{10}b_{10}$ and $a_{14}b_{14}$, which have
been computed previously \cite{2011a}. The representation of figure 4 offers
the advantage of quickly identifying a state at which the switching between
patterns comprising cold and hot spots at the centre happens: it is the
point at which the the line representing the mode in question intersects the
straight line representing the 1D mode. For currents higher than the one
corresponding to the switching, the current density at the centre is lower
than that corresponding to the 1D mode and the pattern comprises a cold spot
at the centre; for lower currents the current density at the centre is
higher than that corresponding to the 1D mode and the pattern comprises a
hot spot at the centre.

A third-generation mode (i.e., a mode which branches from another
multidimensional mode rather than from the 1D mode) is also shown in figure
4; the mode $a_{14,1}b_{14,1}$. This mode branches off from the mode $%
a_{14}b_{14}$ through period-doubling bifurcations: the central spot ceases
being circular and extends both upwards and downwards, thus changing the
period of the mode from $\pi/2$ into $\pi$. Note that third-generation modes
branching off from the mode $a_{10}b_{10}$ which have been computed for
discharges in helium \cite{2013g} and krypton \cite{2014c} branch off
through period-doubling bifurcations as well, however these bifurcations
occur in a different way: in one case, every second spot in the ring
gradually moves from the periphery towards the centre of the cathode (one of
the modes in helium and the mode in krypton); in the other every second spot
of the inner ring gradually moves towards the periphery and merges with a
spot of the outer ring (a mode in helium).

Three third-generation modes bifurcating from the mode $a_{3}b_{3}$,
designated $a_{3,1}b_{3,1}$, $a_{3,2}b_{3,2}$, $a_{3,3}b_{3,3}$, are shown
in figure 5. They branch off and rejoin that branch of the mode $a_{3}b_{3}$
which is associated with a ring spot at the periphery. The mode $%
a_{3,1}b_{3,1}$ is associated with a spot pattern comprising three spots at
the periphery of the cathode. $a_{3,2}b_{3,2}$ and $a_{3,3}b_{3,3}$ are
associated with five and, respectively, six spots at the periphery. Since
neither of the patterns shown in figure 5 comprises a apot at the center,
the coordinates $\left( j_{c},\left\langle j\right\rangle \right) $ would be
inconvenient and the coordinates $\left( j_{e},\left\langle j\right\rangle
\right) $ are used, where $j_{e}$ is the current density at a fixed point on
the periphery of the cathode which coincides with the centre of one of the
spots.

The evolution of the spot patterns associated with the mode $a_{3,2}b_{3,2}$
is shown in figure 6. At state $151.82\unit{V}$, which is positioned near
the bifurcation point $a_{3,2}$, the ring spot is slightly non-uniform in
the azimuthal direction. Further away from $a_{3,2}$, the non-uniformity
gives rise to well-pronounced spots; states $151.81\unit{V}$ and $151.84%
\unit{V}$. The spots become smaller as the current is further reduced; state 
$152.26\unit{V}$. As the bifurcation point $b_{3,2}$ is approached, the
spots expand, state\ $167.94\unit{V}$. In the close vicinity of $b_{3,2}$
(state $170.70\unit{V}$) a ring spot with a small non-uniformity in the
azimuthal direction is formed.

The patterns associated with the mode $a_{3,2}b_{3,2}$ strongly resemble
those observed in \cite{Zhu2014}. (We remind that when neutralization of
charged particles at the wall of the vessel is taken into account in the
modelling, spots at the periphery will be shifted inside the cathode.) The
transition from a spot pattern comprising five spots into a pattern
comprising a ring spot seems to be the same that was found in the modelling
between modes $a_{3,2}b_{3,2}$ and $a_{3}b_{3}$. Note that in the experiment
this transition occurred in a non-stationary way. This may be consistenent
with the presence of the two turning points in the modelling in the vicinity
of the bifurcation point $a_{3,2}$, which may cause a hysteresis.

The behaviour of the modes $a_{3,1}b_{3,1}$ and $a_{3,3}b_{3,3}$ follows the
same trend as the behaviour of the mode $a_{3,2}b_{3,2}$. The patterns are
similar to experimentally observed patterns comprising three and six spots
inside the cathode \cite{Takano2006,Zhu2014}. Note that the pattern with
three spots associated with the mode $a_{3,1}b_{3,1}$ is similar to the
pattern with three spots appearing in some states belonging to $a_{4}b_{4}$
(states $151.15\unit{V}$ and $151.53\unit{V}$ in figure 2b)) and it is
difficult to know which one of these two modes was observed in the
experiments \cite{Takano2006,Zhu2014}.

Third-generation modes bifurcating from the mode $a_{3}b_{3}$ have been
studied also for krypton. The modes which have been computed for krypton
(and not shown for brevity) comprise, in addition to the above-described
modes $a_{3,1}b_{3,1}$, $a_{3,2}b_{3,2}$ and $a_{3,3}b_{3,3}$, also modes
which are associated with patterns comprising nine, twelve, and fifteen
spots at the periphery. Note that the latter modes have not been found in
xenon, which is consistent with patterns in krypton being in general richer
than those in xenon for comparable conditions \cite{2014c}.

\section{Conclusions}

\label{Conclusion}

Seven new 3D modes have been computed for xenon. Three of these modes branch
off from and rejoin the 1D mode; second-generation modes. The other four are
third-generation modes, i.e., bifurcate not from the 1D mode but rather from
another 2D or 3D mode. In the case where the latter mode is 3D as well, the
branching happens through period-doubling bifurcations, similarly to what
was found in the previous modelling for helium and krypton, however the
period-doubling occurs differently. The patterns associated with computed 3D
modes with two, three, four, five and six spots at the periphery of the
cathode and three spots at the peryphery and a spot at the centre are
similar to patterns observed in the experiment. The transition from a spot
pattern comprising five spots into a pattern comprising a ring spot also was
observed in the experiment.

The modelling of self-organization in DC glow discharges has so far been
performed for a cylindrical discharge vessel with parallel-plane electrodes.
Self-organized patterns of spots have been observed in this geometry \cite%
{Takano2006_ICOPS}. However, the most of observations of self-organized
patterns have been performed in cathode boundary layer discharge (CBLD)
devices, which comprise a planar cathode and a ring shaped anode separated
by a ring-shaped dielectric layer. Computation of patterns for the CBLD
configuration is an important topic for future work.

\section{Acknowledgments}

This work was supported by FCT of Portugal through projects \linebreak
PTDC/FIS-PLA/2708/2012 \textit{Modelling, understanding, and controlling
self-organization phenomena in plasma-electrode interaction in gas
discharges: from first principles to applications} and
PEst-OE/MAT/UI0219/2011. D. F. N. Santos is thankful to FCT of Portugal for
the support through the PhD grant SFRH/BD/85068/2012. The authors are
grateful to Dr. WeiDong Zhu for discussion of the experiment \cite{Zhu2014}.

\begin{center}
\begin{tabular}{ll}
\FRAME{itbpFU}{2.6368in}{2.0859in}{0in}{\Qcb{(a)}}{}{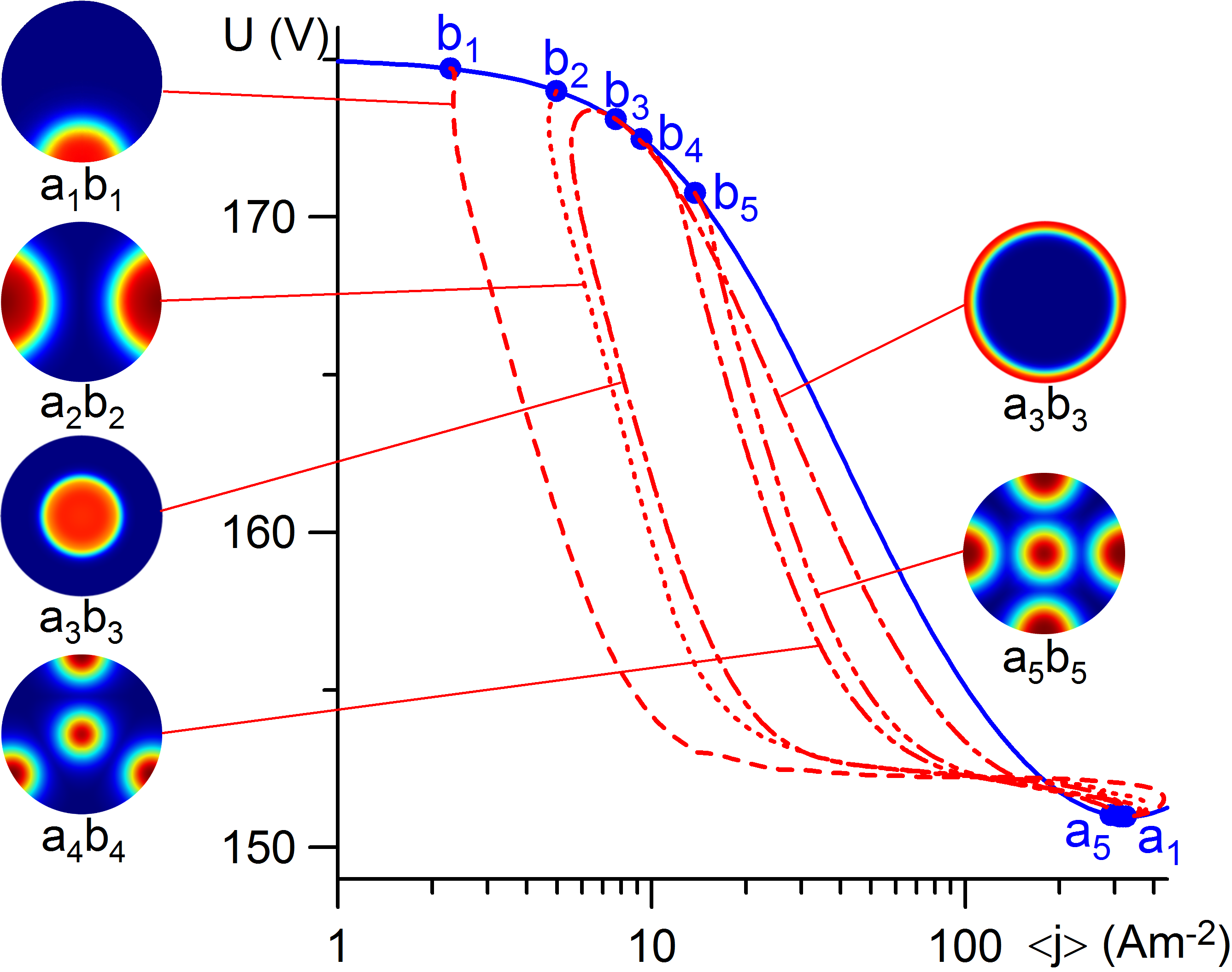}{\special{
language "Scientific Word"; type "GRAPHIC"; maintain-aspect-ratio TRUE;
display "USEDEF"; valid_file "F"; width 2.6368in; height 2.0859in; depth
0in; original-width 8.9154in; original-height 7.0301in; cropleft "0";
croptop "1"; cropright "1"; cropbottom "0"; filename
'fig_01a.png';file-properties "XNPEU";}} & \FRAME{itbpFU}{2.6377in}{2.1292in%
}{0in}{\Qcb{(b)}}{}{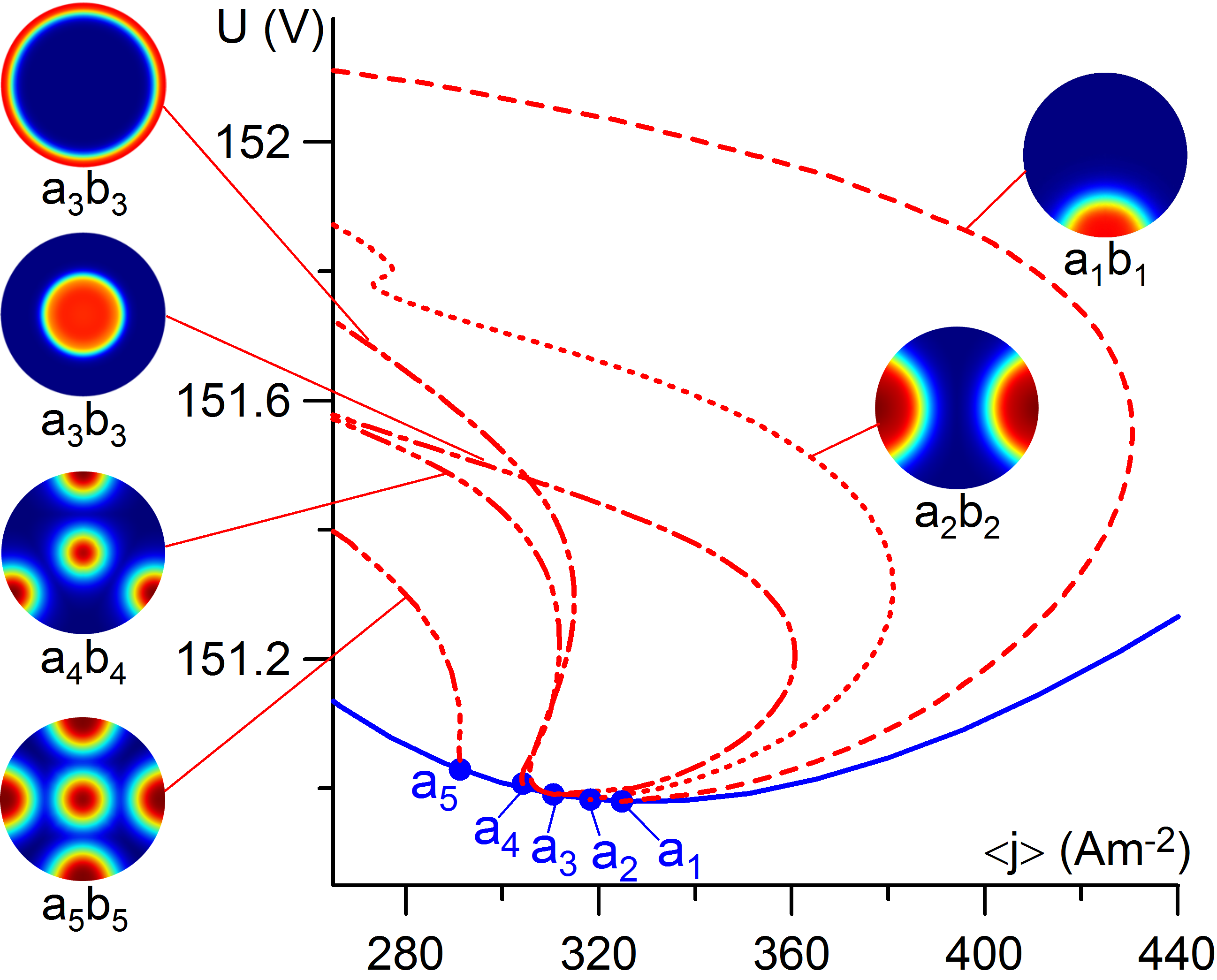}{\special{ language "Scientific Word"; type
"GRAPHIC"; maintain-aspect-ratio TRUE; display "USEDEF"; valid_file "F";
width 2.6377in; height 2.1292in; depth 0in; original-width 8.6464in;
original-height 6.9635in; cropleft "0"; croptop "1"; cropright "1";
cropbottom "0"; filename 'fig_01b.png';file-properties "XNPEU";}}%
\end{tabular}

Figure 1: CVCs. \textrm{Xe}, $30\unit{Torr}$. Solid: the 1D mode.
Dashed-dotted: 2D mode $a_{3}b_{3}$. Other lines: different 3D modes.
Circles: bifurcation points. a): General view. b): Details near the point of
minimum of the CVC of the 1D mode.

\FRAME{dtbpF}{5.2477in}{3.0268in}{0pt}{}{}{_fig2.tif}{\special{language
"Scientific Word";type "GRAPHIC";maintain-aspect-ratio TRUE;display
"USEDEF";valid_file "F";width 5.2477in;height 3.0268in;depth
0pt;original-width 10.2083in;original-height 5.8643in;cropleft "0";croptop
"1";cropright "1";cropbottom "0";filename '_fig2.tif';file-properties
"XNPEU";}}

Figure 2: Evolution of distributions of current on the surface of the
cathode associated with different modes. \textrm{Xe}, $30\unit{Torr}$. a):
mode $a_{2}b_{2}$. b): $a_{4}b_{4}$. c): $a_{5}b_{5}$.

\FRAME{dtbpF}{5.2477in}{5.0903in}{0pt}{}{}{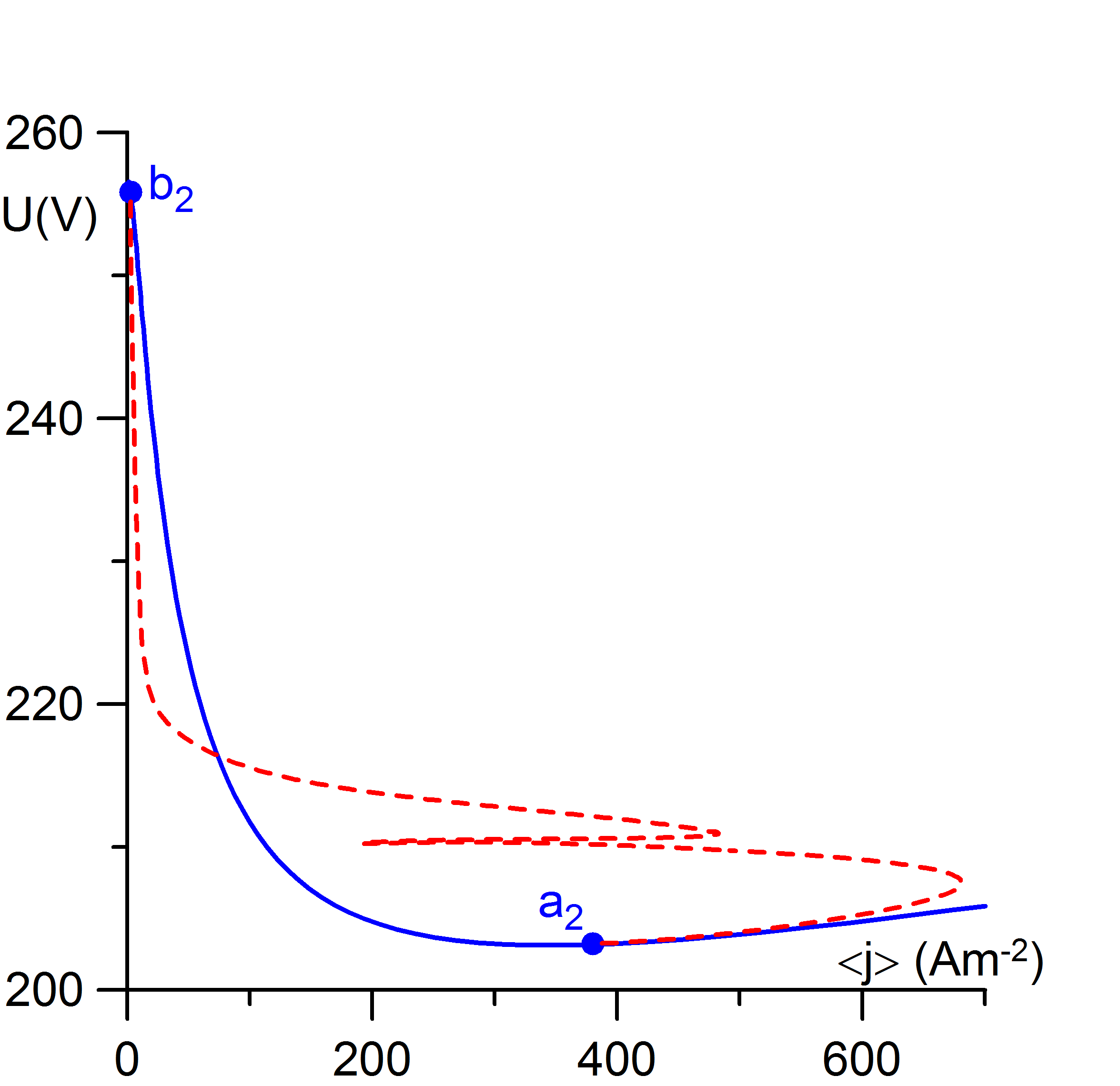}{\special{language
"Scientific Word";type "GRAPHIC";maintain-aspect-ratio TRUE;display
"USEDEF";valid_file "F";width 5.2477in;height 5.0903in;depth
0pt;original-width 7.8369in;original-height 7.6in;cropleft "0";croptop
"1";cropright "1";cropbottom "0";filename 'fig_03.png';file-properties
"XNPEU";}}

Figure 3: CVCs. \textrm{Kr}, $100\unit{Torr}$. Solid: the 1D mode. Dashed:
3D mode $a_{2}b_{2}$. Circles: bifurcation points.

\FRAME{dtbpF}{5.2477in}{5.0332in}{0pt}{}{}{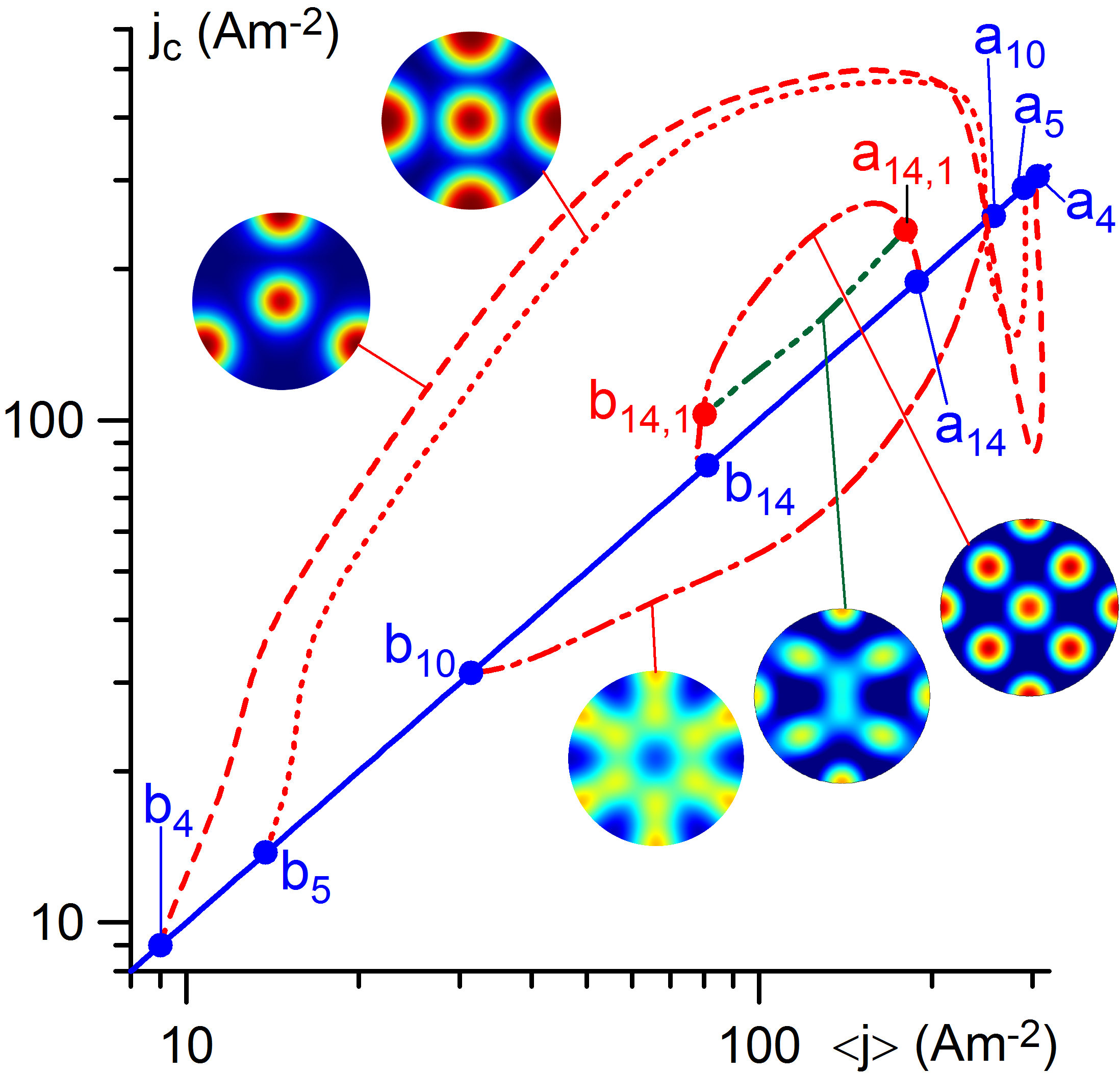}{\special{ language
"Scientific Word"; type "GRAPHIC"; maintain-aspect-ratio TRUE; display
"USEDEF"; valid_file "F"; width 5.2477in; height 5.0332in; depth 0pt;
original-width 7.3206in; original-height 7.0197in; cropleft "0"; croptop
"1"; cropright "1"; cropbottom "0"; filename 'fig_04.png';file-properties
"XNPEU";}}

Figure 4: Bifurcation diagram. \textrm{Xe}, $30\unit{Torr}$. Solid: the 1D
mode. Other lines: 3D modes. Circles: bifurcation points.

\begin{tabular}{ll}
\FRAME{itbpFU}{2.6368in}{2.4223in}{0in}{\Qcb{(a)}}{}{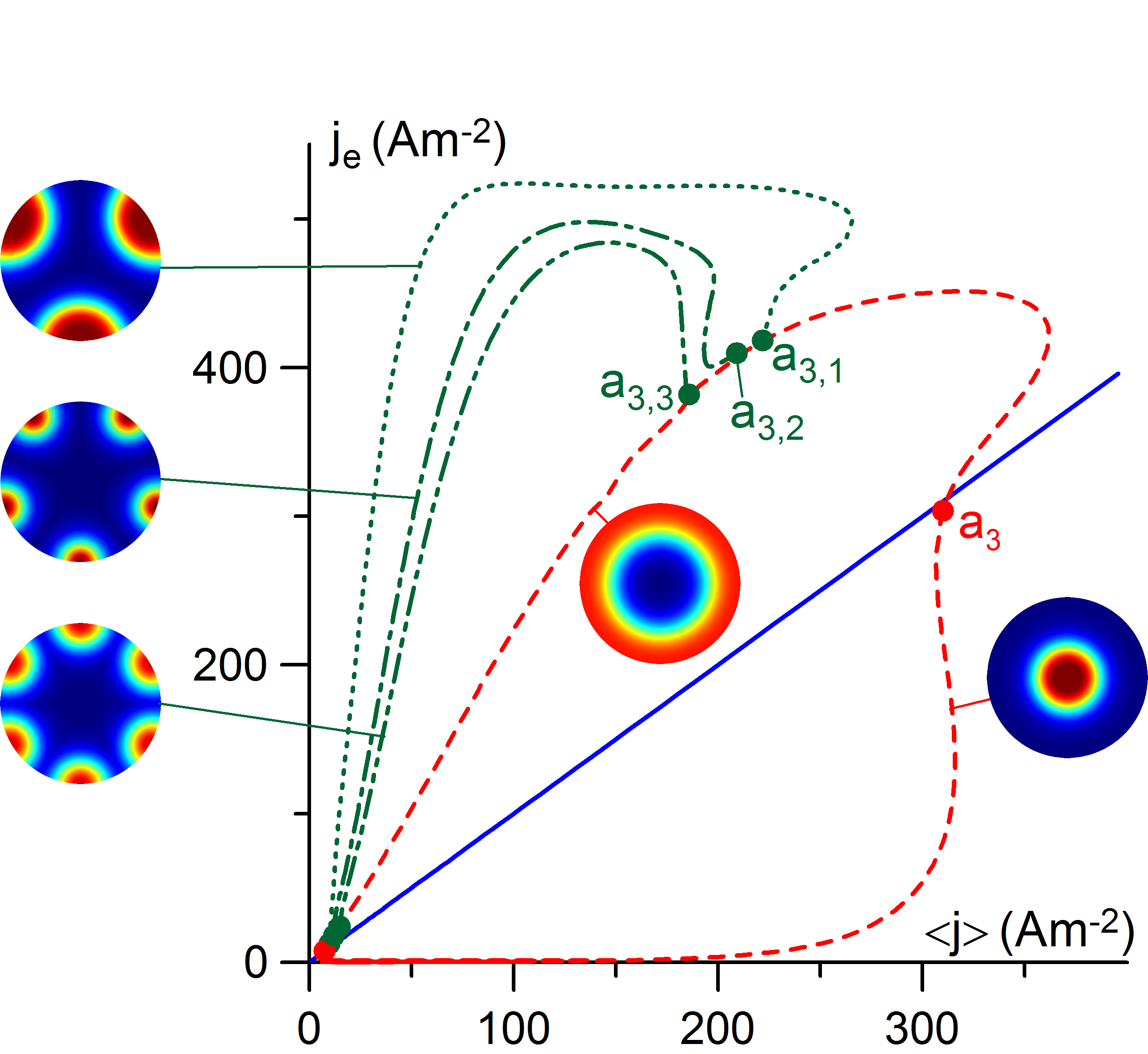}{\special{
language "Scientific Word"; type "GRAPHIC"; maintain-aspect-ratio TRUE;
display "USEDEF"; valid_file "F"; width 2.6368in; height 2.4223in; depth
0in; original-width 8.4267in; original-height 7.7332in; cropleft "0";
croptop "1"; cropright "1"; cropbottom "0"; filename
'fig_05a.png';file-properties "XNPEU";}} & \FRAME{itbpFU}{2.1543in}{2.4267in%
}{0in}{\Qcb{(b)}}{}{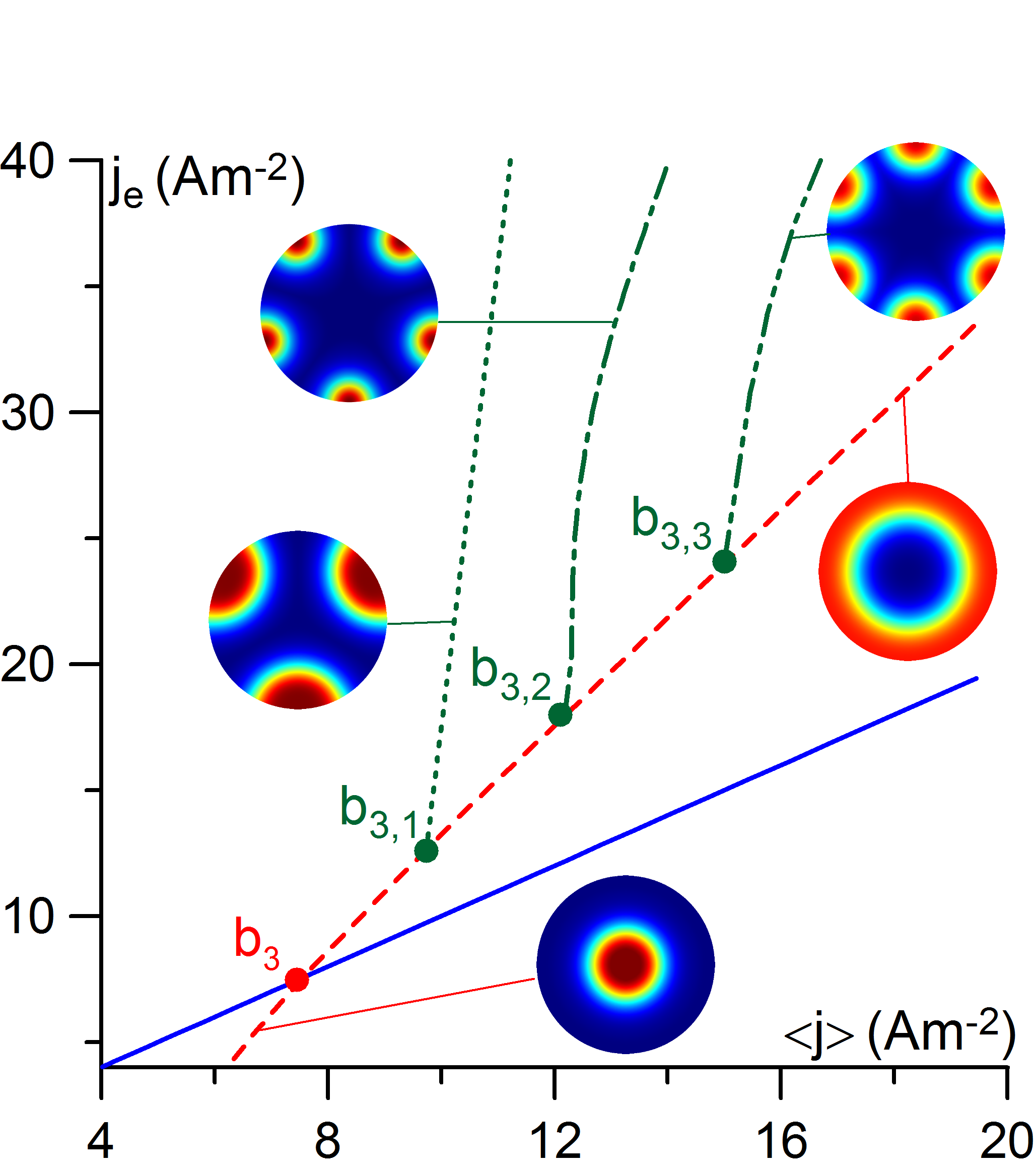}{\special{ language "Scientific Word"; type
"GRAPHIC"; maintain-aspect-ratio TRUE; display "USEDEF"; valid_file "F";
width 2.1543in; height 2.4267in; depth 0in; original-width 6.8571in;
original-height 7.7332in; cropleft "0"; croptop "1"; cropright "1";
cropbottom "0"; filename 'fig_05b.png';file-properties "XNPEU";}}%
\end{tabular}

Figure 5: Bifurcation diagram. \textrm{Xe}, $30\unit{Torr}$. Solid: the 1D
mode. Dashed: 2D mode $a_{3}b_{3}$. Other lines: 3D modes. Circles:
bifurcation points. a): General view. b): Details near the bifurcation point 
$b_{3}$.

\FRAME{dtbpF}{2.2286in}{4.1623in}{0pt}{}{}{fig6.tif}{\special{language
"Scientific Word";type "GRAPHIC";maintain-aspect-ratio TRUE;display
"USEDEF";valid_file "F";width 2.2286in;height 4.1623in;depth
0pt;original-width 2.9473in;original-height 5.5365in;cropleft "0";croptop
"1";cropright "1";cropbottom "0";filename 'fig6.tif';file-properties
"XNPEU";}}

Figure 6: Evolution of distribution of current on the surface of the cathode
associated with the mode $a_{3,2}b_{3,2}$. \textrm{Xe}, $30\unit{Torr}$.
\end{center}

\bibliographystyle{iopart-num}
\bibliography{arc,benilov,Haken1983,mario,pedro,pedro_old}

\end{document}